\documentclass[3p,times,twocolumn]{elsarticle}

\usepackage{ecrc-ac}
\usepackage{amsmath,url,slashed}
\usepackage[figuresright]{rotating}

\newcommand{\sq}{\tilde{q}}
\newcommand{\sqbar}{\bar{\sq}}

\newcommand{\scl}{\tilde{c}_L}

\newcommand{\ssl}{\tilde{s}_L}

\newcommand{\ssr}{\tilde{s}_R}

\newcommand{\neutone}{\tilde{\chi}^0_1}

\newcommand\matO{{\cal O}}

\newcommand{\HWG}{\textsc{Herwig++}}
\newcommand{\DS}{\textsc{Dipole-Shower}}
\newcommand{\PYTH}{\textsc{Pythia}}

\newcommand{\newc}{\newcommand}
\newc{\be}{\begin{equation}}
\newc{\ee}{\end{equation}}
\newc{\bi}{\begin{itemize}}
\newc{\ei}{\end{itemize}}
\newc{\benu}{\begin{enumerate}}
\newc{\eenu}{\end{enumerate}}
\newc{\bc}{\begin{center}}
\newc{\ec}{\end{center}}
\newc{\bfig}{\begin{figure}}
\newc{\efig}{\end{figure}}
\newc{\qbar}{\bar{q}}
\newc{\go}{\tilde{g}}
\newc{\PB}{\textsc{Powheg-Box}}

\newc{\dGa}{\text{d}\Gamma^{\sq_1 \rightarrow \neutone q}}
\newc{\dGb}{\text{d}\Gamma^{\sq_2 \rightarrow \neutone q}}
\newc{\Ga}{\Gamma_{\text{tot}}^{\sq_1}}
\newc{\Gb}{\Gamma_{\text{tot}}^{\sq_2}}
\newc{\GaLO}{\Gamma_{\text{tot},0}^{\sq_1}}
\newc{\GaNLO}{\Gamma_{\text{tot},1}^{\sq_1}}
\newc{\GbLO}{\Gamma_{\text{tot},0}^{\sq_2}}
\newc{\GbNLO}{\Gamma_{\text{tot},1}^{\sq_2}}
\newc{\ds}{\text{d}\sigma}

\volume{00}

\firstpage{1}

\journalname{Nuclear Physics B Proceedings Supplement}

\runauth{M.\ Kr\"amer and M.\ M\"uhlleitner}

\jid{nuphbp}

\jnltitlelogo{Nuclear Physics B Proceedings Supplement}

\usepackage{amssymb}
\biboptions{compress}

\begin{document}

\begin{frontmatter}



\dochead{\small{\flushleft{\vspace*{-25mm}KA-TP-02-2015\\SFB/CPP-14-114\\SLAC-PUB-16200\\TTK-15-04\\[5mm]}}}

\title{Higher-order QCD corrections to supersymmetric particle production and decay at the LHC}


\author[rwth,slac]{Michael Kr\"amer} \author[kit]{Margarete M\"uhlleitner}

\address[rwth]{Institute for Theoretical Particle Physics and Cosmology, RWTH Aachen University, D-52056 Aachen, Germany}
\address[slac]{SLAC National Accelerator Laboratory, Stanford University, Stanford, CA 94025, USA}
\address[kit]{Institute for Theoretical Physics, Karlsruhe Institute of Technology, 76128 Karlsruhe, Germany}

\begin{abstract}
We review recent results on higher-order calculations to squark and gluino production and decay at the LHC, as obtained within the 
Collaborative Research Centre / Transregio 9 ``Computational Particle Physics``. In particular, we 
discuss inclusive cross sections, including the summation of threshold corrections, higher-order calculations 
for specific squark production channels and for top squark decays, and next-to-leading order calculations for exclusive observables matched 
to parton showers. 
\end{abstract}


\end{frontmatter}

%%%%%%%%%%%%%%%%%%%%%%%%%%%%%%%%%%%%%%%%%%%%%%%%%%%%%%%%%%%%%%%%
\section{Introduction}
\label{sec:introduction}
%%%%%%%%%%%%%%%%%%%%%%%%%%%%%%%%%%%%%%%%%%%%%%%%%%%%%%%%%%%%%%%%

After the discovery of a Higgs particle in the first phase of the LHC
in 2012, the key scientific goal for LHC run\,2 starting in
2015 is to search for and explore physics beyond the Standard Model
(SM). Supersymmetric (SUSY) theories~\cite{Wess:1974tw} are among the
most attractive extensions of the Standard Model. They allow for the
unification of the electromagnetic, weak and strong gauge
couplings~\cite{Langacker:1991an,Amaldi:1991cn,Ellis:1990wk} and
provide a solution to the fine-tuning problem of the
SM~\cite{Gildener:1976ai,Veltman:1980mj}, if some of the
supersymmetric particles have masses near the TeV
scale~\cite{Barbieri:1987fn}. Furthermore, in supersymmetric models
with $R$-parity conservation~\cite{Fayet:1977yc,Farrar:1978xj}, the
lightest supersymmetric particle (LSP) is stable and may constitute
the dark matter in the universe~\cite{Goldberg:1983nd,Ellis:1983ew}.

A generic LHC signature for supersymmetric models with $R$-parity
conservation are cascade decays of the strongly interacting
supersymmetric particles, squarks and gluinos, which terminate in a
weakly interacting LSP and thus result in missing energy signatures.
Current LHC searches for supersymmetry in jets plus missing energy
final states place lower limits on the masses of squarks and gluinos
at around 1.5\,TeV~\cite{Aad:2014bva, Chatrchyan:2014lfa}.  Note,
however, that these limits are not generic for supersymmetry, but
depend on certain assumptions about the supersymmetry breaking and the
resulting SUSY particle mass spectrum.

Given the importance of SUSY searches at the LHC, accurate theoretical
predictions for the production and decay of supersymmetric particles
are required. The precision calculations are crucial to derive
accurate limits on SUSY masses and couplings, or to determine the
properties of supersymmetric particles in the case of discovery.

In this contribution we shall review the calculation of higher-order
QCD corrections to squark and gluino production and decay at the
LHC. In Section~\ref{sec:production} we first present results for the
inclusive production cross sections, including the summation of large
logarithmic corrections. The
higher-order QCD corrections to SUSY particle decays are discussed in
Section~\ref{sec:decay}. Fully differential next-to-leading order 
predictions for squark production and decay, including the
effects of parton showers, are  presented in
Section~\ref{sec:production_and_decay}. We conclude in
Section~\ref{sec:conclusions}.

%%%%%%%%%%%%%%%%%%%%%%%%%%%%%%%%%%%%%%%%%%%%%%%%%%%%%%%%%%%%%%%%
\section{Squark and gluino production at the LHC}
\label{sec:production}
%%%%%%%%%%%%%%%%%%%%%%%%%%%%%%%%%%%%%%%%%%%%%%%%%%%%%%%%%%%%%%%%

We consider the minimal supersymmetric extension of the Standard Model
(MSSM)~\cite{Nilles:1983ge,Haber:1984rc} where, as a consequence of
$R$-parity conservation, squarks and gluinos are pair-produced in
proton-proton collisions at the LHC:
\begin{equation}
  pp \;\to\; \tilde{q}\tilde{q}\,,
  \tilde{q}\bar{\tilde{q}}\,, \tilde{q}\tilde{g}\,, \tilde{g}\tilde{g},\, 
  \tilde t_1\bar{\tilde t}_1,\, \tilde t_2\bar{\tilde t}_2 + X\,.
\label{eq:processes}
\end{equation}
The production of top squarks (stops), $ \tilde t_{1,2}$, has to be
treated separately, because the strong Yukawa coupling between top
quarks, stops and Higgs fields gives rise to potentially large mixing
effects and mass splitting~\cite{Ellis:1983ed}. In
Eq.~(\ref{eq:processes}) and throughout the rest of this paper,
$\tilde t_1$ and $\tilde t_2$ denote the lighter and heavier stop mass
eigenstate, respectively. For the other squarks we suppress the
chiralities, i.e.\ $\tilde{q} =(\tilde{q}_{L}, \tilde{q}_{R})$, and do
not explicitly state the charge-conjugated processes.

The cross sections for the hadro-production of squarks and gluinos in
the MSSM are known including next-to-leading order (NLO)
QCD~\cite{Beenakker:1996ch, Beenakker:1997ut, Beenakker:1999xh,
  Berger:2000iu, Berger:2002vd} and
electroweak~\cite{Bornhauser:2007bf,Hollik:2007wf,
  Beccari2008mi,Hollik:2008yi, Hollik:2008vm,
  Mirabell2009ap, Arhrib:2009sb,Germer:2010vn,
  Germer:2011an,Germer:2014jpa} corrections. The QCD corrections
are particularly significant and included in the public computer code
{\sc Prospino}~\cite{Beenakker:1996ch, Beenakker:1997ut, Beenakker:1999xh, Beenakker:1996ed}. A large part of the QCD corrections can be attributed
to the emission of soft gluons and can be taken into account to all
orders in perturbation theory by means of threshold resummation
techniques at next-to-leading logarithmic (NLL) or next-to-next-to-leading logarithmic (NNLL)
accuracy~\cite{Kulesz2008jb, Langenfeld:2009eg,
  Kulesz2009kq, Beneke:2009rj, Beenakker:2009ha,
  Beenakker:2010nq, Beneke:2010da, Beenakker:2011fu,
  Kauth:2011vg, Kauth:2011bz, Beenakker:2011sf,
  Falgari:2012hx, Langenfeld:2012ti, Falgari:2012sq,
  Pfoh:2013iia, Broggio:2013uba, Beenakker:2013mva,
  Beenakker:2014sma}.  NLO+NLL predictions for the production of
strongly interacting MSSM particles as implemented in the computer
code {\sc Nll-fast}~\cite{Kulesz2008jb, Kulesz2009kq,
  Beenakker:2009ha, Beenakker:2010nq, Beenakker:2011fu} are
currently state-of-the-art, and are employed by the LHC experiments
and by large parts of the theory community to interpret search limits
and constrain the MSSM parameter space~\cite{Kramer:2012bx,
  Borschensky:2014cia}.

The calculations implemented in \textsc{Prospino} and
\textsc{Nll-fast} assume that five flavors of left- and right-chiral
squarks, $\tilde{u}_{L,R}$, $\tilde{d}_{L,R}$, $\tilde{c}_{L,R}$,
$\tilde{s}_{L,R}$, and $\tilde{b}_{L,R}$, are mass-degenerate. NLO-QCD
predictions for generic MSSM spectra have been presented in the
literature recently~\cite{Hollik:2012rc,GoncalvesNetto:2012yt,
  Gavin:2013kga, Hollik:2013xwa, Gavin:2014yga}, and we shall comment
on those in Section~\ref{sec:non_degenerate}. Moreover, with the
\textsc{MadGolem}~\cite{
  GoncalvesNetto:2012yt,Binoth:2011xi,Goncalves:2014axa} and \textsc{MadGraph5\_aMC@NLO}~\cite{Alwall:2014hca}
  frameworks there exist automated tools for the calculation of NLO-QCD corrections to generic
 supersymmetric processes at the LHC.
  
While the NLO-QCD corrections for degenerate squarks have been
available for many years~\cite{Beenakker:1996ch,
  Beenakker:1997ut}, the summation of large logarithmic threshold
corrections and the extension of the NLO calculations to generic MSSM
spectra have been achieved more recently. We will thus focus on
threshold resummation and the impact of non-degenerate MSSM particle
spectra on the QCD corrections. The summation of the
next-to-leading logarithmic corrections and the corresponding tool
{\sc Nll-fast} are presented in Section~\ref{sec:nllfast}. In
Section~\ref{sec:nnll} we comment on the recent resummation of NNLL
terms, while the NLO-QCD effects for generic MSSM spectra are
discussed in Section~\ref{sec:non_degenerate}.

% ----------------------------------------------------------------------------------------------------------------------------------------------------------------------------
\subsection{NLL threshold resummation} \label{sec:nllfast}
% ----------------------------------------------------------------------------------------------------------------------------------------------------------------------------

A significant part of the NLO-QCD corrections to heavy particle
production at the LHC can be attributed to the kinematic region where
the partonic center-of-mass energy is close to the production
threshold. Near threshold, the NLO corrections are typically large,
with the most significant contributions coming from soft-gluon
emission off the colored particles in the initial and final state.
The contributions due to soft gluon emission can be taken into account
to all orders by means of threshold resummation.  Threshold
resummation for MSSM squark and gluino pair-production processes at
NNL accuracy is described below, following
the work presented in Refs.~\cite{Kulesz2008jb, Kulesz2009kq,
  Beenakker:2009ha, Beenakker:2010nq, Beenakker:2011fu}. We
shall comment on NNLL resummation in Section~\ref{sec:nnll}.

The resummation for $(2\to 2)$ QCD processes has been studied
extensively in the literature, specifically for
heavy-quark~\cite{Kidonakis:1997gm,Bonciani:1998vc} and jet
production~\cite{Kidonakis:1998bk, Kidonakis:1998nf, Bonciani:2003nt}.
The calculations presented in Refs.~\cite{Kulesz2008jb,
  Kulesz2009kq, Beenakker:2009ha, Beenakker:2010nq,
  Beenakker:2011fu} make use of the framework developed there.

The hadronic threshold for the inclusive production of two final-state
particles $k, l$ with masses $m_k$ and $m_l$ corresponds to a hadronic
centre-of-mass energy squared that is equal to $S=(m_k+m_l)^2$.  Thus
we define the threshold variable $\rho$, measuring the distance from
threshold in terms of energy fraction, as
\begin{equation}
  \label{eq:1}
\rho \;=\; \frac{(m_k+m_l)^2}{S}\,.  
\end{equation}
The numerical results presented below are based on the following
expression for the NLL-resummed cross section, matched to the exact
NLO calculation~\cite{Beenakker:1996ch, Beenakker:1997ut}:
\begin{eqnarray}
\label{eq:14}
\hspace*{-6mm}\sigma^{\rm NLO+NLL}_{pp \to kl}\bigl(\rho, \{m^2\},\mu^2\bigr) \!\!\!\!
  &=&\!\!\!\! \sigma^{\rm NLO}_{pp \to kl}\bigl(\rho, \{m^2\},\mu^2\bigr)\nonumber
          \\[1mm]
   &&  \hspace*{-39mm}+\, \frac{1}{2 \pi i}\! \sum_{i,j=q,\bar{q},g}\! \int_\mathrm{CT}\!\!dN\,\rho^{-N}\,
       \tilde f_{i/p}(N+1,\mu^2)\,\tilde f_{j/p}(N+1,\mu^2) \nonumber\\[1mm]
   && \hspace*{-38mm} \times\,
       \left[\tilde\sigma^{\rm res}_{ij\to kl}\bigl(N,\{m^2\},\mu^2\bigr)
             \,-\, \tilde\sigma^{\rm res}_{ij\to kl}\bigl(N,\{m^2\},\mu^2\bigr)
       {\left.\right|}_{\scriptscriptstyle {\rm NLO}}\, \right]. 
\end{eqnarray}
The $\tilde f_{i/p}, \tilde f_{j/p}$ are the parton distribution functions in Mellin space,  
and the last term in the square brackets denotes the NLL resummed
expression expanded to NLO.  $\mu$ is the common factorization and
renormalization scale. The resummation is performed after taking a
Mellin transform (indicated by a tilde) of the cross section,
\begin{equation}
  \label{eq:10}
  \tilde\sigma_{pp \to kl}\bigl(N, \{m^2\}\bigr) 
 \equiv \int_0^1 d\rho\;\rho^{N-1}\;
           \sigma_{pp\to kl}\bigl(\rho,\{ m^2\}\bigr) \,.
\end{equation}
To evaluate the contour CT of the inverse Mellin transform in
Eq.~(\ref{eq:14}) the so-called ``minimal
prescription''~\cite{Catani:1996yz} has been adopted.  The NLL
resummed cross section in Eq.~(\ref{eq:14}) reads
\begin{eqnarray}
  \label{eq:12}
\lefteqn{\tilde{\sigma}^{\rm res} _{ij\rightarrow
    kl}\bigl(N,\{m^2\},\mu^2\bigr)=} \nonumber\\[1mm]
&&\hspace*{-4mm}\sum_{I}\,\tilde\sigma^{(0)}_{ij\rightarrow kl,I}\bigl(N,\{m^2\},\mu^2\bigr)\, 
      C_{ij \rightarrow kl, I}\bigl(N,\{m^2\},\mu^2\bigr) \nonumber\\[1mm]
&& \hspace*{-4mm}  \times\,\Delta_i (N+1,Q^2,\mu^2)\,\Delta_j (N+1,Q^2,\mu^2)\nonumber\\[1mm]
 &&  \hspace*{-4mm}  \times\, \Delta^{\rm (s)}_{ij\rightarrow
       kl,I}\bigl(N+1,Q^2,\mu^2\bigr)\,,
\end{eqnarray}
where $\tilde{\sigma}^{(0)}_{ij \rightarrow kl, I}$ are the
color-decomposed leading-order cross sections in Mellin-moment space,
with $I$ labelling the possible color structures.  Here the hard
scale $Q^2 = (m_k + m_l)^2$ has been introduced.  The perturbative
functions $C_{ij \rightarrow kl, I}$ contain information about hard
contributions beyond leading order.  This information is only relevant
beyond NLL accuracy and therefore $C_{ij \rightarrow kl,I} =1$ is used
in the calculations. The functions $\Delta_{i}$ and $\Delta_{j}$ sum
the effects of the (soft-)collinear radiation from the incoming
partons. They are process-independent and do not depend on the color
structures.  They contain the leading logarithmic dependence, as well
as part of the subleading logarithmic behaviour. The expressions for
$\Delta_{i}$ and $\Delta_{j}$ can be found in the
literature~\cite{Kulesz2009kq}. The resummation of the soft-gluon
contributions, which depends on the color structures in which the
final state SUSY particle pairs can be produced, contributes at the
NLL level and is summarized by the factor
\begin{equation}
  \Delta_{I}^{\rm (s)}\bigl(N,Q^2,\mu^2\bigr) 
  \;=\; \exp\left[\int_{\mu}^{Q/N}\frac{dq}{q}\,\frac{\alpha_{\rm s}(q)}{\pi}
                 \,D_{I} \,\right]\,.
\label{eq:2}
\end{equation}
The one-loop coefficients $D_{I}$ follow from the threshold limit of
the one-loop soft anomalous-dimension
matrix~\cite{Kulesz2009kq,Beenakker:2009ha}.

\begin{figure}[t!]
\begin{center}
\includegraphics[width=0.475\textwidth]{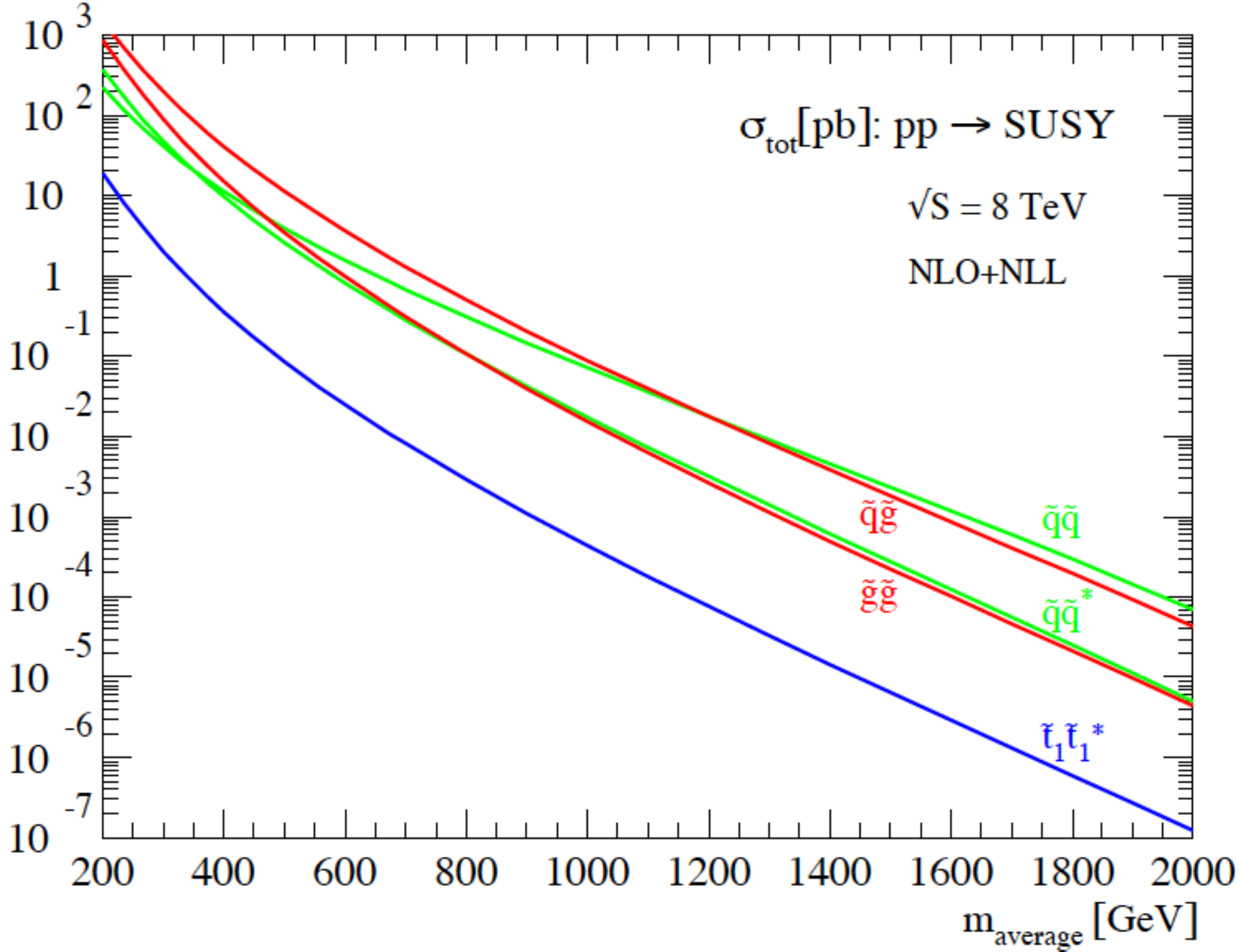}
\includegraphics[width=0.475\textwidth]{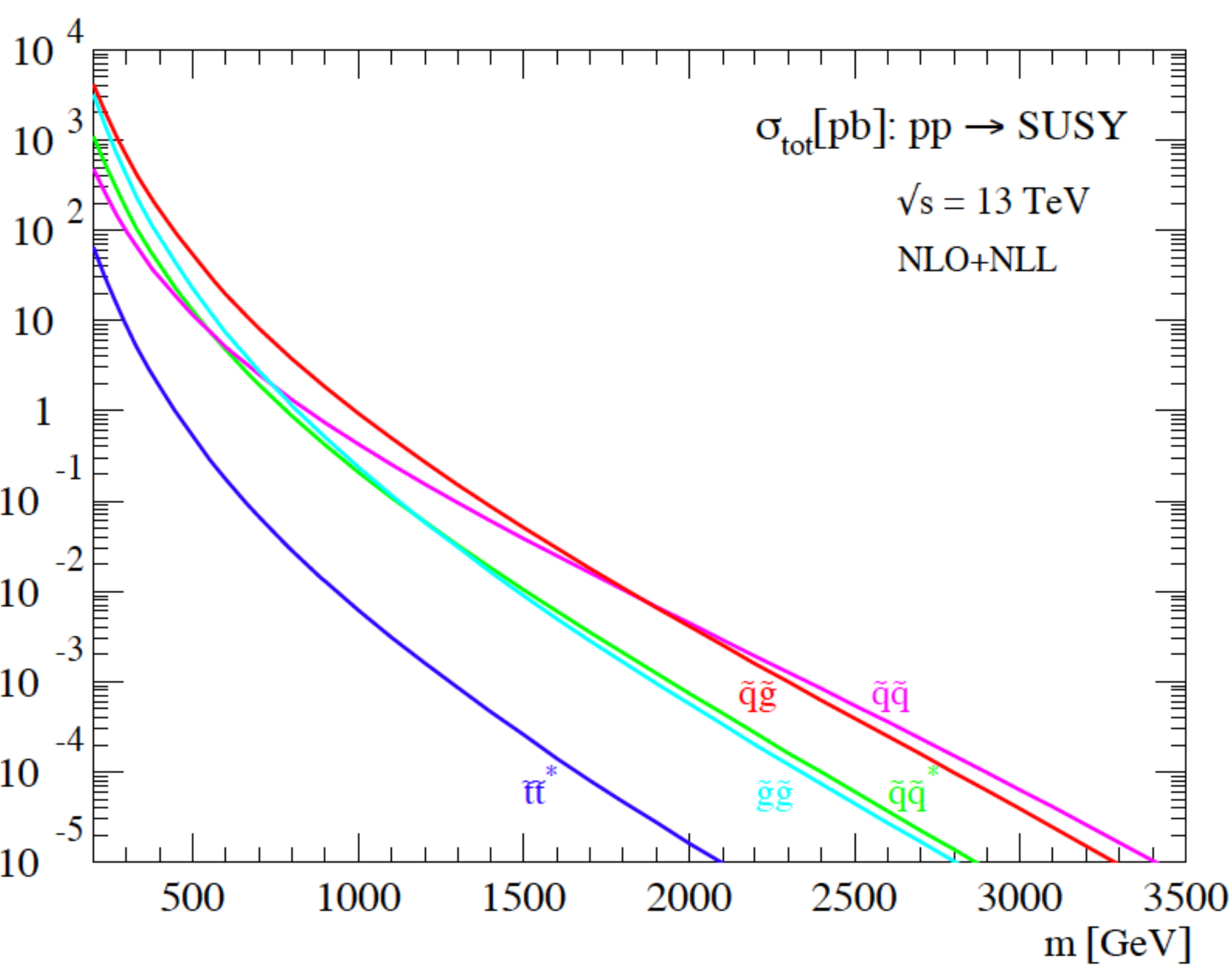}
\end{center}
\vspace*{-5mm}
\caption{\it \label{fig:nlo_nll_all} NLO+NLL cross sections for squark
  and gluino production at the LHC ($\sqrt{S}=8$ and 13\,TeV). (The $\tilde{t}\tilde{t}^*$ curve in the lower plot corresponds to the 
  $\tilde{t}_1\tilde{t}^*_1$ cross section; both notations $\tilde{q}^*$ and $\bar{\tilde{q}}$ are used in the literature to denote antisquarks.) 
  The
  cross sections have been evaluated for a common squark and gluino
  mass $m_{\tilde{q}} = m_{\tilde{g}} = m_{\rm average} = m$, and the renormalization
  and factorization scales have been set to $\mu = m$. The MSTW2008
  pdfs \cite{Martin:2009iq} have been used.}
\end{figure}

In Fig.~\ref{fig:nlo_nll_all} we present the NLO+NLL predictions for
squark and gluino production at the LHC ($\sqrt{S}=8$ and $13$\,TeV).
The $\overline{\rm MS}$-scheme with five active flavors is used to
define $\alpha_{\rm s}$ and the parton distribution functions (pdfs) at
NLO. The masses of the squarks and gluinos are renormalized in the
on-shell scheme, and the SUSY particles are decoupled from the running
of $\alpha_{\rm s}$ and the pdf. The results have been obtained with the 
MSTW2008 parton distribution function~\cite{Martin:2009iq}. 

We observe that the inclusive cross section is dominated by
$\tilde{q}\tilde{q}$ and $\tilde{q}\tilde{g}$ production at large
sparticle masses, a consequence of the large valence quark component
of the pdf at large $x$. Note that the $\tilde{q}\tilde{q}$,
$\tilde{q}\tilde{g}$ and $\tilde{q}\bar{\tilde{q}}$ cross sections include
the sum over five flavors of squarks, $\tilde{q} \in \{\tilde{u},
\tilde{d}, \tilde{c}, \tilde{s}, \tilde{b}\}$, and both chiralities
($\tilde{q}_L$ and $\tilde{q}_R$). Therefore, and because of a
different threshold behaviour, the cross section for the production of
light stops, $pp\to \tilde{t}_1\bar{\tilde{t}}_1$, is strongly
suppressed.

\begin{figure}
\begin{center}
\includegraphics[width=0.475\textwidth]{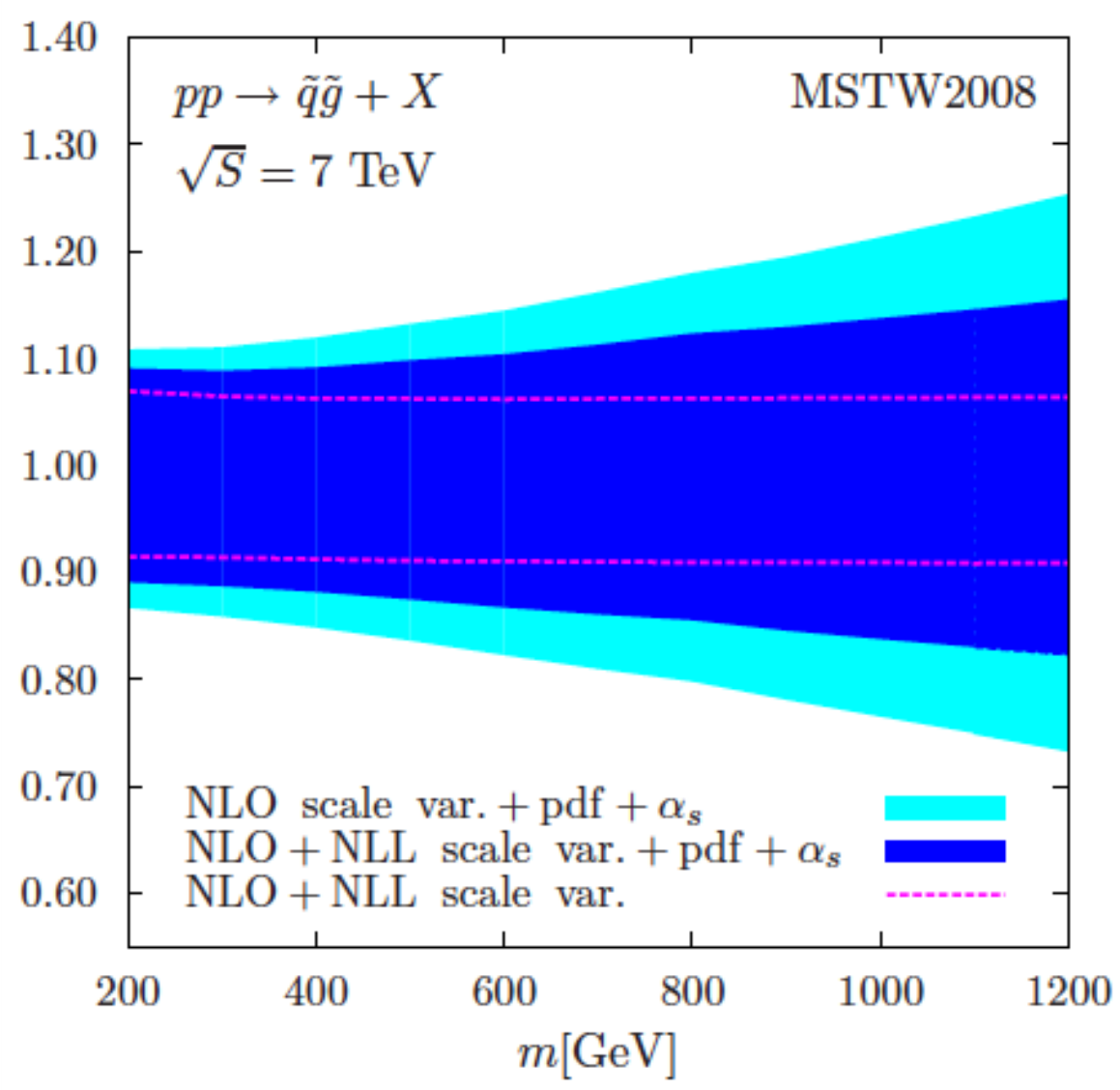}
\end{center}
\vspace*{-5mm}
\caption{\it \label{fig:nll_error} The theoretical uncertainty for
  associated squark-gluino production at the LHC with 7\,TeV, $pp \to
  \tilde{q}\tilde{g}\,+X$, as a function of the sparticle mass
  $m_{\tilde{q}} = m_{\tilde{g}} = m$. The error bands represent the
  NLO+NLL scale uncertainty in the range $m/2\le \mu \le 2m$, and
  the total theory uncertainty including the 68\% C.L.\ pdf and
  $\alpha_{\rm s}$ error, added in quadrature, and the error from
  scale variation in the range $m/2\le \mu \le 2m$ added linearly to
  the combined pdf and $\alpha_{\rm s}$ uncertainty. The total theory
  uncertainty is shown at both NLO and NLO+NLL. The MSTW2008
  pdfs \cite{Martin:2009iq} have been used.}
  \end{figure}

The NLL summation increases the cross section prediction if the
renormalization and factorization scales are chosen near the average
mass of the final state particles. More importantly, threshold
resummation leads to a significant reduction of the scale dependence
over the full range of sparticle masses, with an overall scale
uncertainty at NLO+NLL of less than 10\%. This is shown in
Fig.\,\ref{fig:nll_error} for squark-gluino associated production. We
also show the full theory uncertainty, consisting of the 68\% C.L.\
pdf and $\alpha_{\rm s}$ error added in quadrature, combined linearly
with the scale variation error for the NLO cross sections and for the
NLO+NLL cross sections.  We find that even though the pdf uncertainty
is significant, the inclusion of threshold resummation leads to a
sizeable reduction of the overall theory uncertainty.

  The NLO+NLL predictions for degenerate squark masses can be computed
  with the program {\sc Nll-fast}~\cite{nll_fast} and are employed by
  the LHC experiments to
  interpret search limits and constrain the MSSM parameter
  space. Selected results for squark and gluino cross sections at
  current and future hadron colliders are collected in
  Refs.~\cite{Kramer:2012bx, Borschensky:2014cia}, including a
  more detailed discussion of the theoretical scale and pdf uncertainty.

% ----------------------------------------------------------------------------------------------------------------------------------------------------------------------------
\subsection{NNLL threshold resummation } \label{sec:nnll}
% ----------------------------------------------------------------------------------------------------------------------------------------------------------------------------

Let us first briefly comment on the ingredients needed to extend the
resummation of threshold logarithms to next-to-next-to-leading logarithmic 
accuracy. As discussed in Section~\ref{sec:nllfast}, the
(soft-)collinear radiation effects are described by the functions
$\Delta_i$ and $\Delta_j$, while wide-angle soft radiation is
accounted for by $\Delta^{\rm (s)}_{ij\to kl,I}$ in Eq.~(\ref{eq:12}). The
product of these radiative factors can be written schematically as
\begin{eqnarray}
  \label{eq:NNLL-expa}
\lefteqn{\hspace*{-5mm}\Delta_i\Delta_j\Delta^{\rm(s)}_{ij\to kl,I}
  =}\nonumber \\ &&\hspace*{-5mm}\exp\Big[L g_1(\alpha_{\rm s}L) + g_2(\alpha_{\rm s}L) + \alpha_{\rm s}g_3(\alpha_{\rm s}L) + \ldots \Big]  \,.
\end{eqnarray}
This exponent contains all the dependence on large logarithms $L=\ln
N$. The leading logarithmic approximation (LL) is represented by the
$g_1$ term alone, whereas the NLL approximation requires additionally
including the $g_2$ term. Similarly, the $g_3$ term is needed for the
NNLL approximation. The expressions for the $g_1$ and $g_2$
functions can be found in e.g.~\cite{Kulesz2009kq} and that for
the NNLL $g_3$ function in e.g.~\cite{Beenakker:2011sf}.

The matching coefficients $C_{ij\to kl,I}$ in~Eq.~(\ref{eq:12}) collect
non-logarithmic terms as well as logarithmic terms of non-soft origin
in the Mellin moments of the higher-order contributions. The
coefficients $C_{ij\to kl,I}$ factorise into a part that contains the
Coulomb corrections and a part containing hard
contributions~\cite{Beneke:2010da}
\begin{eqnarray}
\lefteqn{C_{ij\to kl,I}= }\nonumber\\
&&(1+ \frac{\alpha_{\rm s}}{\pi}{\cal C}_{ij\to kl,I}^{\rm Coul,(1)}+\frac{\alpha_{\rm s}^2}{\pi^2}{\cal C}_{ij\to kl,I}^{\rm Coul,(2)}+\dots)\nonumber\\
&\times&(1+ \frac{\alpha_{\rm s}}{\pi}{\cal C}_{ij\to kl,I}^{(1)}+ \frac{\alpha_{\rm s}^2}{\pi^2}{\cal C}_{ij\to kl,I}^{(2)}+\dots)\,.
\label{eq:factCcoeff}
\end{eqnarray}
Apart from the terms of ${\cal O}(\alpha_s)$, which need to be
included in $C_{ij\to kl,I}$ when performing resummation at NNLL, some
of the ${\cal O}(\alpha_s^2)$ terms are also known and can be included
in the numerical calculations. Expanding Eq.~(\ref{eq:factCcoeff}) we have
\begin{eqnarray}
\lefteqn{\hspace*{-5mm}C^{\rm NNLL}_{ij\to kl,I}=}\nonumber\\
&&\hspace*{-7mm}1+\frac{\alpha_{\rm s}}{\pi}\left({\cal C}^{\rm Coul,(1)}_{ij\to kl,I}+{\cal C}^{\rm (1)}_{ij\to kl,I}\right)\nonumber\\
&&\hspace*{-4mm}+\frac{\alpha_{\rm s}^2}{\pi^2}\left({\cal C}^{\rm Coul,(2)}_{ij\to kl,I}+{\cal C}^{\rm (2)}_{ij\to kl,I}+ {\cal C}^{\rm (1)}_{ij\to kl,I}{\cal C}^{\rm Coul,(1)}_{ij\to kl,I}\right).\label{eq:matchingcoeff}
\end{eqnarray} 
The first-order hard matching coefficients ${\cal C}^{\rm (1)}_{ij\to
  kl,I}$ were calculated in~\cite{Beenakker:2013mva}, whereas the
expressions for the first-order Coulomb corrections ${\cal C}^{\rm
  Coul,(1)}_{ij\to kl,I}$ in Mellin-moment space are listed
in~\cite{Beenakker:2014sma}. The form of the two-loop Coulomb
corrections in $x$-space is known in the
literature~\cite{Beneke:2009ye}, and the ${\cal C}^{\rm
  Coul,(2)}_{ij\to kl,I}$ coefficients have been calculated
in~\cite{Beenakker:2014sma} by taking Mellin moments of the
near-threshold approximation of these two-loop Coulomb corrections.
The second-order hard matching coefficient ${\cal C}^{\rm (2)}_{ij\to
  kl,I}$ is not known at the moment and we put ${\cal C}^{\rm
  (2)}_{ij\to kl,I}=0$ in~Eq.~(\ref{eq:matchingcoeff}).

Once we have the NNLL resummed cross section in Mellin-moment space,
we match it to the approximated NNLO cross section, which is
constructed by adding the near-threshold approximation of the NNLO
correction~\cite{Beneke:2009ye} to the full NLO
result~\cite{Beenakker:1996ch}. The matching is performed according
to
\begin{eqnarray}
  \label{eq:matching}
 \lefteqn{\hspace*{-5mm}\sigma^{\rm (NNLL~matched)}_{h_1 h_2 \to kl}\bigl(\rho, \{m^2\},\mu^2\bigr) 
  = \sigma^{\rm (NNLO_{Approx})}_{h_1 h_2 \to kl}\bigl(\rho, \{m^2\},\mu^2\bigr)}
          \nonumber\\
&& \hspace*{-10mm}+\, \sum_{i,j}\,\int_\mathrm{CT}\,\frac{dN}{2\pi i}\,\rho^{-N}\,
       \tilde f_{i/h_1}(N+1,\mu^2)\,\tilde f_{j/h_{2}}(N+1,\mu^2) \nonumber\\
&&\hspace*{-5mm}\times\,
       \left[\tilde\sigma^{\rm(res,NNLL)}_{ij\to kl}\bigl(N,\{m^2\},\mu^2\bigr)\right.\nonumber\\
      &&   \left.    \,-\, \tilde\sigma^{\rm(res,NNLL)}_{ij\to kl}\bigl(N,\{m^2\},\mu^2\bigr)
       {\left.\right|}_{\scriptscriptstyle{\rm (NNLO_{Approx})}}\, \right]. 
\end{eqnarray}
To evaluate the inverse Mellin transform in~Eq.~(\ref{eq:matching}) we
again adopt the prescription of reference~\cite{Catani:1996yz}
for the integration contour CT.

\begin{figure}
\begin{center}
\includegraphics[width=0.475\textwidth]{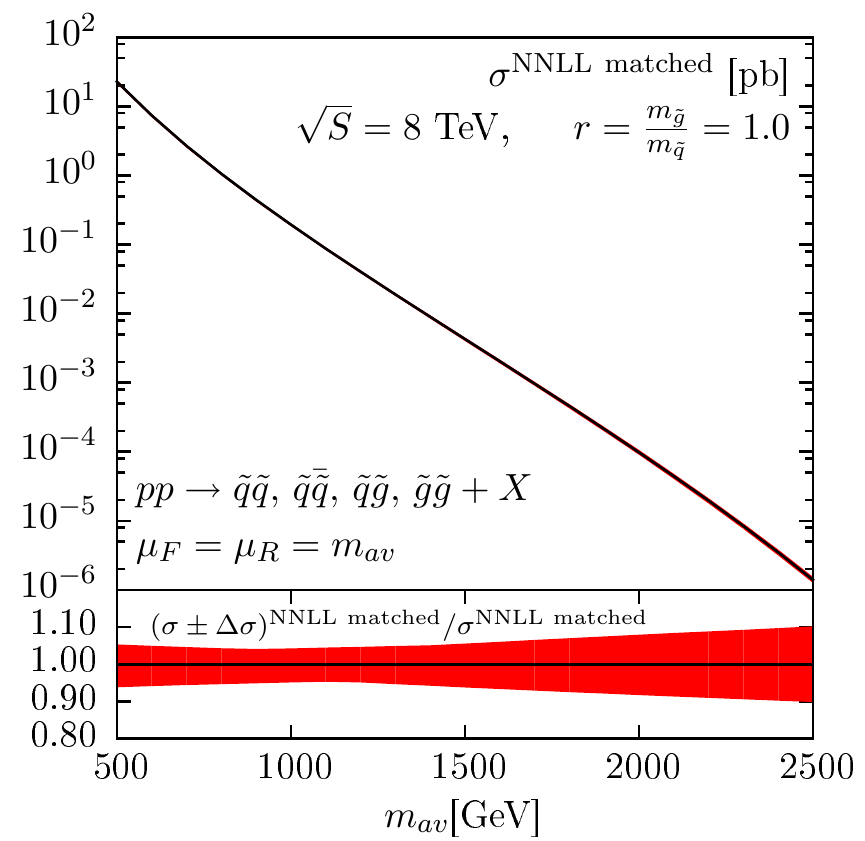}
\end{center}
\vspace*{-5mm}
\caption{\it \label{fig:nnll_error} The NNLL matched cross section for
  the sum of the four processes of pair production of squarks and
  gluinos, including the theoretical error. The error band includes
  the 68\% C.L. pdf and $\alpha_{\rm s}$ errors, added quadratically,
  and the scale uncertainty varied in the range $m_{av}/2 \leq \mu
  \leq 2m_{av}$, added linearly to the combined pdf and $\alpha_{\rm
    s}$ error. The energy is that of the LHC at 8 TeV. The squark and
  gluino masses have been taken equal and the common renormalization
  and factorization scale has been set equal to the average mass of
  the two particles produced. The MSTW 2008 NNLO parton distribution
  function~\cite{Martin:2009iq} has been used.}
\vspace*{-3mm}
\end{figure}

In Fig.~\ref{fig:nnll_error} we present the NNLL matched cross section
prediction for the sum of the different squark and gluino production
processes. The theoretical uncertainty includes the scale error as
well as pdf and $\alpha_{\rm s}$ errors. It is obtained by linearly
adding the scale dependence in the range $m/2 \leq \mu \leq 2m$ to the
combined 68\% C.L. pdf and $\alpha_{\rm s}$ uncertainties, the latter
two added in quadrature.  We see uncertainties grow from approximately
5\% for masses near the present lower bounds, to 10\% for masses
approaching 2.5 TeV. Including the NNLL contributions leads to a
further reduction of the scale dependence for the squark and gluino production
total cross sections, 
with the exception of gluino-pair production as discussed in more detail in Ref.\,\cite{Beenakker:2014sma}. 

% ----------------------------------------------------------------------------------------------------------------------------------------------------------------------------
\subsection{NLO-QCD corrections for generic MSSM spectra} \label{sec:non_degenerate}
% ----------------------------------------------------------------------------------------------------------------------------------------------------------------------------

The calculations implemented in \textsc{Prospino} and
\textsc{Nll-fast} described in the previous section assume that five
flavors of left- and right-chiral squarks, $\tilde{u}_{L,R}$,
$\tilde{d}_{L,R}$, $\tilde{c}_{L,R}$, $\tilde{s}_{L,R}$, and
$\tilde{b}_{L,R}$, are mass-degenerate. NLO-QCD predictions for
generic MSSM spectra have been presented in the literature
recently~\cite{Hollik:2012rc,GoncalvesNetto:2012yt, Gavin:2013kga,
  Hollik:2013xwa, Gavin:2014yga}, and we shall briefly comment on
these calculations below.

We shall follow Ref.\,\cite{Gavin:2014yga} and discuss the production
of squark-antisquark pairs as an example.  General NLO squark and
gluino production cross sections have also been presented
in~\cite{GoncalvesNetto:2012yt}.  At LO squark-antisquark
production, $pp \to \tilde{q}\bar{\tilde{q}}$, proceeds through
quark-antiquark annihilation and gluon-fusion,
\begin{equation}
\begin{aligned}
 q_i\, \qbar_j &\rightarrow \sq_k^{\,c1}\, \sqbar_l^{\,c2}\, ,\\ 
g\, g &\rightarrow \sq_i^{\,c}\, \sqbar_i^{\,c}\, .
\end{aligned}
\label{eq:borncontri}
\end{equation}
Here, the lower indices indicate the flavor of the particle, whereas
the upper indices for the squarks denote the chirality.
Due to the flavor conserving structure of the relevant vertices the
$gg$ initiated processes and the $s$-channel gluon exchange only contribute to
the production of squarks of the same flavor and chirality. We shall
focus on the production of squarks of the first two generations.  In
total, this includes 64 possible final state combinations. This number
can be reduced to 36 independent channels if the invariance under
charge conjugation is taken into account.

To illustrate the size of the SUSY-QCD corrections for the individual
production channels, let us consider a particular benchmark scenario
of the constrained MSSM~ \cite{AbdusSalam:2011fc}, specified by universal GUT-scale
parameters $m_0 / m_{1/2} / A_0 = 825 / 550 / 0$\,GeV, $\tan(\beta) =
10$ and $\text{sgn}(\mu) = +1$. The corresponding squark and gluino
masses read \renewcommand{\arraystretch}{1.25} \bc{\small
\begin{tabular}{c|c|c|c|c }
  $m_{\tilde{u}_L} = m_{\scl}$\!\! & \!\!$m_{\tilde{u}_R} = m_{\tilde{c}_R}$ \!\!\! &\!\!\!  $m_{\tilde{d}_L} = m_{\ssl}$ \!\!\!\!\! &\!\!\!  $m_{\tilde{d}_R} = m_{\ssr}$ \!\!\! &\!\! \!\! $m_{\tilde{g}}\!\!\!\!$ \\[1mm]\hline
  $\!\! 1799.53\!\! $ & $\!\! 1760.21\!\! $  & $\!\! 1801.08\!\! $ & $\!\! 1756.40\!\! $ & $\!\! 1602.96\!\!$ \\[1mm]
\end{tabular}}
\ec
%\vspace*{-0.5cm}
where we have neglected the very small mass splitting between the
first and second generation squarks.

In Table~\ref{tab:totxs} we present LO and NLO results for individual
squark-antisquark production channels at the LHC with 14\,TeV.  The
renormalization and factorization scales have been set to the average
squark mass, i.e.\ $\overline{m}_{\sq} =1779.31\,\text{GeV}$, and the
\textsc{CTEQ6L1}~\cite{Pumplin:2002vw} and
\textsc{CT10NLO}~\cite{Lai:2010vv} parton distribution functions have
been adopted for the LO and NLO cross sections, respectively.

 \begin{table}[t]
\renewcommand{\arraystretch}{1.25}
\bc
\begin{tabular}{c | c |c  | c }
Process  &  $\sigma_{\text{LO}} [\text{fb}]$  &  $\sigma_{\text{NLO}} [\text{fb}]$  &  K-factor \\\hline
$\tilde{u}_{L}\bar{\tilde{u}}_{L}$  & $  9.51\cdot 10^{-2} $ & $ 1.43\cdot 10^{-1} $ & $ 1.50 $ \\ 
$\tilde{u}_{R}\bar{\tilde{u}}_{R}$  & $  1.14\cdot 10^{-1} $ & $ 1.72\cdot 10^{-1} $ & $ 1.51 $ \\ 
$\tilde{d}_{L}\bar{\tilde{d}}_{L}$  & $  5.50\cdot 10^{-2} $ & $ 8.79\cdot 10^{-2} $ & $ 1.60 $ \\ 
$\tilde{d}_{R}\bar{\tilde{d}}_{R}$  & $  6.89\cdot 10^{-2} $ & $ 1.11\cdot 10^{-1} $ & $ 1.61 $ \\ 
$\tilde{u}_{L}\bar{\tilde{u}}_{R}$  & $  3.75\cdot 10^{-1} $ & $ 5.12\cdot 10^{-1} $ & $ 1.37 $ \\ 
$\tilde{d}_{L}\bar{\tilde{d}}_{R}$  & $  1.41\cdot 10^{-1} $ & $ 1.70\cdot 10^{-1} $ & $ 1.21 $ \\ 
$\tilde{u}_{L}\bar{\tilde{d}}_{L}$  & $  6.98\cdot 10^{-2} $ & $ 7.89\cdot 10^{-2} $ & $ 1.13 $ \\ 
$\tilde{u}_{L}\bar{\tilde{d}}_{R}$  & $  2.98\cdot 10^{-1} $ & $ 3.54\cdot 10^{-1} $ & $ 1.19 $ \\ 
$\tilde{u}_{R}\bar{\tilde{d}}_{L}$  & $  2.94\cdot 10^{-1} $ & $ 3.49\cdot 10^{-1} $ & $ 1.19 $ \\ 
$\tilde{u}_{R}\bar{\tilde{d}}_{R}$  & $  8.36\cdot 10^{-2} $ & $ 9.54\cdot 10^{-2} $ & $ 1.14 $ \\ 
\hline
Sum  & 1.59 & 2.07 & 1.30   \\
\end{tabular}
\caption{\label{tab:totxs}The LO and NLO cross sections for squark-antisquark production of the first generation at the LHC with $\sqrt{S} = 14$\,TeV
obtained for a specific scenario of the constrained MSSM as described in the text. The
renormalization and factorization scales have been set to the average
squark mass, i.e.\ $\overline{m}_{\sq} =1779.31\,\text{GeV}$, and the
\textsc{CTEQ6L1}~\cite{Pumplin:2002vw} and
\textsc{CT10NLO}~\cite{Lai:2010vv} parton distribution functions have
been adopted for the LO and NLO cross sections, respectively.}
\ec
\vspace*{-0.2cm}
\end{table}

Note, that the channels with squarks of the same flavor and chirality
in the final state, displayed in the first four rows of the table,
have contributions from $gg$ initial states and therefore larger
$K$-factors than channels with squarks of different flavor or
chirality. Hence, approximating the individual $K$-factors by the
average $K$-factor as obtained from \textsc{Prospino}, $K_{\rm avg} =
1.39$, is not very accurate .

Determining the individual corrections consistently is especially
important if the decays are taken into account and the branching
ratios of the different squarks differ significantly for the specific
decay channel under consideration. In order to assess the possible
numerical impact of this approximation we consider the decay
$\sq\rightarrow q \neutone$ at LO at the level of total cross
sections, i.e.\ we multiply the production cross sections for the
individual squark-antisquark channels with the respective
LO branching ratios:
\begin{eqnarray}
\lefteqn{\hspace*{-5mm}\sum_{\text{channels}}\sigma_{\text{NLO}}\cdot \text{BR}^{\text{LO}}\left(\sq\rightarrow \neutone q \right) \cdot \text{BR}^{\text{LO}}\left(\sqbar\rightarrow \neutone \qbar \right)}\nonumber \\&&= 0.139\, \text{fb}.
\end{eqnarray}

Multiplying the LO result for each subchannel with the average
$K$-factor, $K_{\rm avg} = 1.39$, and the corresponding branching
ratios gives
\begin{eqnarray}
\lefteqn{\hspace*{-5mm}\sum_{\text{channels}}\sigma_{\text{LO}}\cdot K^{\text{avg}}\cdot \text{BR}^{\text{LO}}\left(\sq\rightarrow \neutone q \right) \cdot \text{BR}^{\text{LO}}\left(\sqbar\rightarrow \neutone \qbar \right)}\nonumber\\
 && =0.126\, \text{fb}\, .
\end{eqnarray}
Thus the rate obtained with the approximation relying on an average 
$K$-factor for all subchannels is roughly $10\%$ smaller for this
particular case.

To summarize this section, we have seen that using an average
$K$-factor to globally rescale individual sub-channels for
squark-antisquark production may not be a very accurate
approximation. For a generic benchmark scenario, we find differences
of $\mathcal{O}(10\%)$, i.e.\ of phenomenological relevance for the experimental accuracy 
expected in run 2 of the LHC. In general, the more the branching ratios of
the different squarks differ, the more important the consistent
treatment of the individual corrections becomes.

%%%%%%%%%%%%%%%%%%%%%%%%%%%%%%%%%%%%%%%%%%%%%%%%%%%%%%%%%%%%%%%%
\section{Light stop decays}
\label{sec:decay}
%%%%%%%%%%%%%%%%%%%%%%%%%%%%%%%%%%%%%%%%%%%%%%%%%%%%%%%%%%%%%%%%

Higher-order QCD corrections to supersymmetric particle decay widths and branching ratios have been calculated 
for many years, see e.g.\ Ref.\,\cite{susyhit,Djouadi:2006bz} and references therein. Here, we shall focus on the decays of light stops, 
which play a special role in supersymmetric models. 

In most SUSY models a light stop arises naturally, as the mixing is
proportional to the large top Yukawa coupling, which leads to a large mass
splitting between the stop mass eigenstates. 
Light stops play an important role in view of the Higgs mass and
naturalness arguments; together with the top loops they provide the
dominant higher-order corrections to the light CP-even Higgs mass \cite{Okada:1990vk, higgs_mass2, higgs_mass3, higgs_mass4, higgs_mass5},
which are necessary to shift the tree-level mass to the observed value of about 125~GeV. 
A light stop can also lead to the correct dark matter relic density through
co-annihilation, in particular for mass differences between the stop
and the lightest neutralino $\tilde{\chi}_1^0$ of 15-30~GeV
%, or apseudoscalar mass $M_A$ with $M_A\approx 2 m_{\tilde{\chi}_1^0}$
\cite{Boehm:1999bj, Ellis:2001nx, Balazs:2004bu, Balazs:2004ae, Ellis:2014ipa, deSimone:2014pda}. Moreover, light stops
allow for successful baryogenesis within the MSSM
\cite{Caren1996wj, Caren1997ki, de Carlos:1997ru, Huet:1995sh, Delepine:1996vn, Losad1998at, Losad1999tf, 
Cirigliano:2006dg, Li:2008ez, Caren2008rt, Caren2008vj, Cirigliano:2009yd, Laine:2012jy}. 

While the canonical LHC searches for jets and missing transverse energy place lower limits on the masses of the squarks of the first two
generations of approximately 1.5~TeV~\cite{Aad:2014bva, Chatrchyan:2014lfa}, the lightest stop can 
still be rather light and have a mass below the kinematical threshold
for the decay into a top quark and the lightest neutralino
$\tilde{\chi}_1^0$. If we assume the lightest stop $\tilde{t}_1$ to be the next-to-lightest SUSY particle (NLSP)  
and the lightest neutralino to be the lightest
SUSY particle, then $\tilde{t}_1$ can decay
into the LSP and a charm quark $c$ or an up quark $u$, i.e.~$\tilde{t}_1 \to (u/c)\, \tilde{\chi}_1^0$
\cite{Hikas1987db,Muhlleitner:2011ww}. Another possible decay
channel is the four-body decay $\tilde{t}_1 \to \tilde{\chi}_1^0 b f
\bar{f}'$ \cite{Boehm:1999tr}, where $f$ and $f'$ denote generic light
fermions. The two-body decay into charm/up and neutralino is
flavor-violating (FV). While the MSSM in general exhibits many
sources of flavor violation, so that the decay can already occur at
tree-level, high precision tests in the sector of quark flavor
violation and limits on flavor-violating neutral currents from $K$,
$D$ and $B$ meson studies put stringent constraints on the amount of
possible flavor violation. The hypothesis of Minimal Flavor
Violation (MFV) \cite{mfv1,Hall:1990ac, mfv2,mfv3,mfv4} 
has been proposed in order to solve this New Physics
Flavor Puzzle. It requires that all sources of flavor and
CP-violation are given by the SM structure of the Yukawa couplings, so
that flavor mixing in models of New Physics is then always
proportional to the off-diagonal elements of the
Cabibbo-Kobayashi-Maskawa (CKM) matrix \cite{ckmmatrix,ckm2}. 
However, the hypothesis of MFV is not renormalization group invariant
\cite{mfv3}. Flavor off-diagonal squark mass terms are induced
through the Yukawa couplings, so that the squark and quark mass
matrices cannot be diagonalized simultaneously any more and e.g.\ the stop
state receives some admixture from the charm (up) squark, inducing a 
tree-level flavor-changing neutral current (FCNC) coupling between the 
stop, the lightest neutralino and the charm (up) quark. Note that for very small FV 
stop-neutralino-up/charm quark couplings, the four-body
decay can become important and has to be considered for a
reliable prediction of the $\tilde{t}_1$ branching ratios.
The stop masses have been bounded by LEP
\cite{Abbiendi:1999yz,Abbiendi:2002mp} and Tevatron
\cite{Abazov:2008rc,Aaltonen:2012tq}, and more recently by ATLAS \cite{atlas1} and
CMS \cite{cms1}, with the strongest limits coming from the ATLAS
analyses Refs.~\cite{Aad:2014nra,Aad:2014kra}.  All these
analyses assume a branching ratio equal to one for the considered decay channel
of the $\tilde{t}_1$, i.e.\ either the FV two-body or the
four-body decay. The branching ratios, however, can deviate
significantly from one in large parts of the allowed parameter range,
as has been shown in Refs.~\cite{Muhlleitner:2011ww,Grober:2014aha}
and will be outlined in the following. Taking this effect into account, the
experimental exclusion limits on the stop, which are based on the
assumption of branching ratios equal to one, are considerably
weakened. 

In Ref.\,\cite{Muhlleitner:2011ww} we improved upon the existing approximate
result for the $\tilde{t}_1 \to (u/c) \,\tilde{\chi}_1^0$ decay of
Ref.\,\cite{Hikas1987db}, which only takes into account the leading
logarithms of the MFV scale. We calculated the exact one-loop decay width in the
framework of MFV, by performing the full renormalization program and
including the finite non-logarithmic terms arising from the loop
integrals. For the numerical analysis we investigated two constrained MSSM
scenarios with universal soft-breaking terms at the GUT scale $M_{\rm GUT} \approx
10^{16}$~GeV. This scale is identified with the MFV scale, and all
soft SUSY breaking parameters are family universal. The boundary
conditions at $\mu_{\mbox{\scriptsize MFV}} =M_{\rm GUT}$ are 
\begin{eqnarray}
\begin{array}{lll}
\hspace*{-5mm}(1) & \hspace*{-18mm} m_0 = 200 \mbox{ GeV}; &\hspace*{-10mm} m_{1/2} = 230 \mbox{ GeV}; \\ 
\hspace*{-5mm} A_0 = -920\mbox{ GeV}; &  \tan\beta = 10; & \mbox{sign}(\mu) = + ;\\[2mm]
\hspace*{-5mm}(2) &  \hspace*{-18mm} m_0 = 200 \mbox{ GeV}; & \hspace*{-10mm} m_{1/2} = 230 \mbox{ GeV}; \\ 
\hspace*{-5mm} A_0 = -895 \mbox{ GeV}; & \tan\beta = 10; & \mbox{sign}(\mu) = + . \\
\end{array}
\label{eq:scenarios}
\end{eqnarray}
The mass spectra and mixing angles have been calculated with the
spectrum calculator {\sc SPheno} \cite{Porod:2003um,spheno2}, which allows for
flavor violation. The second scenario
has a larger $\tilde{t}_1-\tilde{\chi}_1^0$ mass difference compared
to scenario (1), whereas the mass difference between $\tilde{t}_1$ and
the lightest chargino $\tilde{\chi}_1^+$ is smaller:
\begin{eqnarray}
\begin{array}{llll}
\hspace*{-0mm}(1) &  m_{\tilde{t}_1} /  m_{\tilde{\chi}_1^0} /  m_{\tilde{\chi}_1^+}  = 104/92/175 \mbox{ GeV}; \\
\hspace*{-0mm}(2) &  m_{\tilde{t}_1} /  m_{\tilde{\chi}_1^0} /  m_{\tilde{\chi}_1^+}  = 130/92/175 \mbox{ GeV} .
\end{array}
\label{eq:scen12}
\end{eqnarray}
The partial stop decay width into charm and neutralino has been
calculated with the full one-loop formula and compared to 
the approximate result \cite{Hikas1987db}. For the latter, the
charged $W$ boson mass $M_W$ has been taken as generic loop 
particle mass. The widths are given in
Table~\ref{tab:partialwidths}. 
\begin{table}[t]
\begin{center}
$
\renewcommand{\arraystretch}{1.25}
\begin{array}{c|c|c}
\tilde{t}_1 \to c \tilde{\chi}_1^0 & \Gamma^{\mbox{\scriptsize{1-loop}}}
\mbox{[GeV]} & \Gamma^{\mbox{\scriptsize{H/K}}}
\mbox{[GeV]}  \\ \hline
\mbox{Scenario}\,(1) & 9.322 \cdot 10^{-10} & 1.004 \cdot
10^{-9} \\ \hline
\mbox{Scenario}\, (2) & 5.862 \cdot 10^{-9} & 6.446 \cdot
10^{-9} \\
\end{array}
$
\caption{The partial widths 
  for the decay $\tilde{t}_1\to  c \tilde{\chi}_1^0$ in two MFV
  scenarios, calculated with the exact 1-loop formula,
  $\Gamma^{\mbox{\scriptsize{1-loop}}}$, 
  and with the approximate formula of Ref.\,\cite{Hikas1987db},
  $\Gamma^{\mbox{\scriptsize{H/K}}}$. \label{tab:partialwidths}}  
\end{center}
\vspace*{-0.5cm}
\end{table}
They have been obtained with the 
program \textsc{Susy-Hit} \cite{susyhit,Djouadi:2006bz}, where the full one-loop formula for
the flavor changing stop decay has been implemented. The results for
the exact and the approximate decay widths differ by ${\cal
  O}(10\%)$. 
In the calculation of the branching ratios the 
$\tilde{t}_1 \to u \tilde{\chi}_1^0$ and the four-body decay widths have
also been included in the total width. The branching
ratios are summarised in Table~\ref{tab:brs}. 
In scenario (2) the larger mass difference between $\tilde{t}_1$ and
$\tilde{\chi}_1^0$ leads to a more important four-body decay width, as it
is dominated by the chargino exchange diagrams
\cite{Boehm:1999tr}. This in turn induces a deviation of the branching
ratio $\mbox{BR}(\tilde{t}_1 \to \tilde{\chi}_1^0 c)$ from one by a few
per-cent, as can be inferred from Table~\ref{tab:brs}.
%%%%%%%%%%%%%%
\begin{table}[h]
\vspace*{0.3cm}
\begin{center}
$
\renewcommand{\arraystretch}{1.25}
\begin{array}{c|c|c} 
\mbox{BR} & \mbox{BR}(\tilde{t}_1 \to \tilde{\chi}_1^0 c)
& \mbox{BR}(\tilde{t}_1 \to \tilde{\chi}_1^0 b f \bar{f}') \\ \hline
\mbox{Scenario}\, (1) & 0.9944  & 4.587 \cdot 10^{-5}\\ \hline
\mbox{Scenario}\, (2) & 0.9443  & 0.0504 
\end{array}
$
\caption{\label{tab:brs} The $\tilde{t}_1$ branching ratios for
  different final states for scenario (1) and (2). For details, see
  \cite{Muhlleitner:2011ww}.}
\end{center}
\vspace*{-0.2cm}
\end{table} 
%%%%%%%%%%%%%%
\begin{figure*}
\begin{center}
\includegraphics[width=8cm]{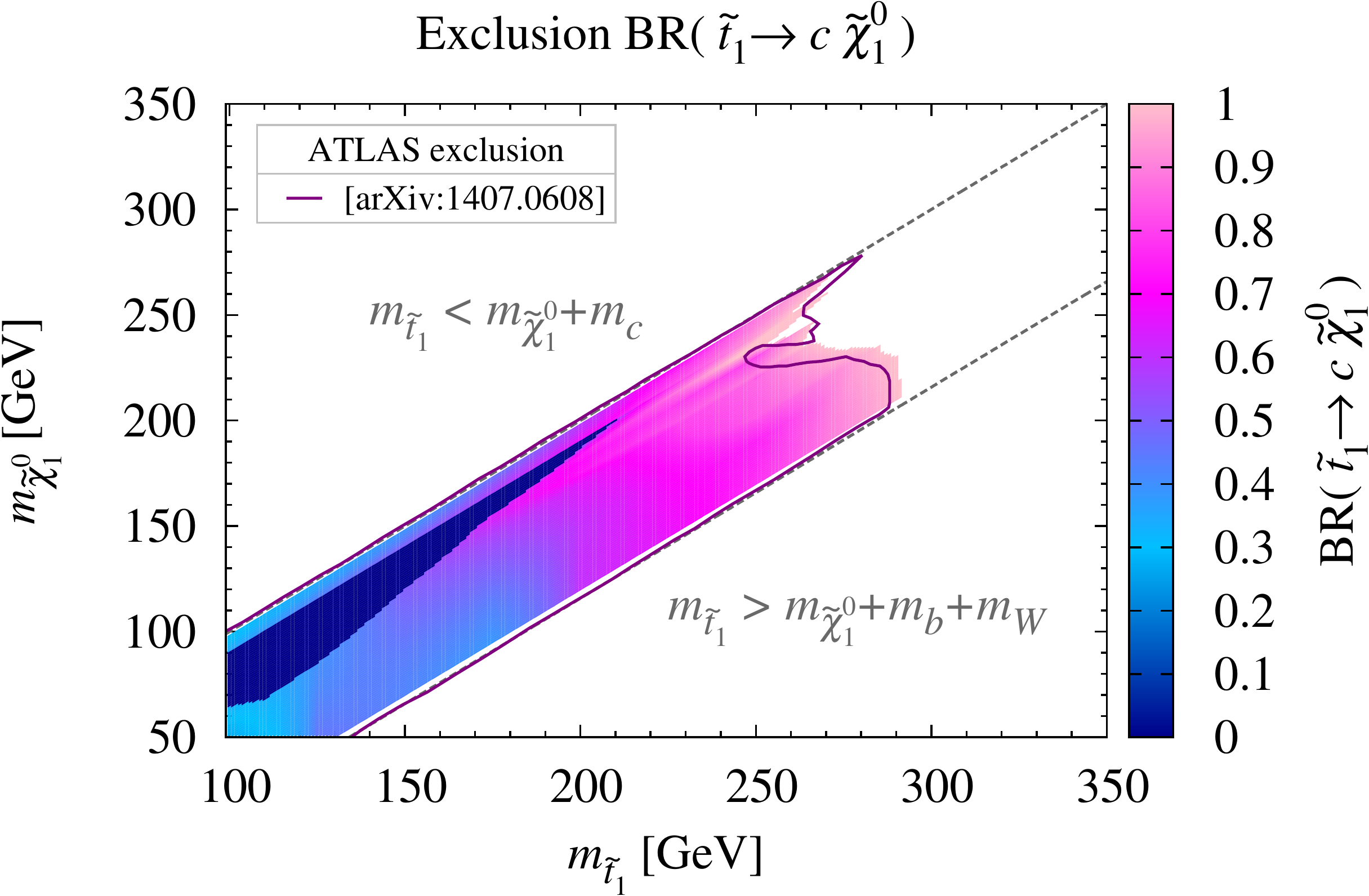} \hspace*{0.2cm}
\includegraphics[width=8cm]{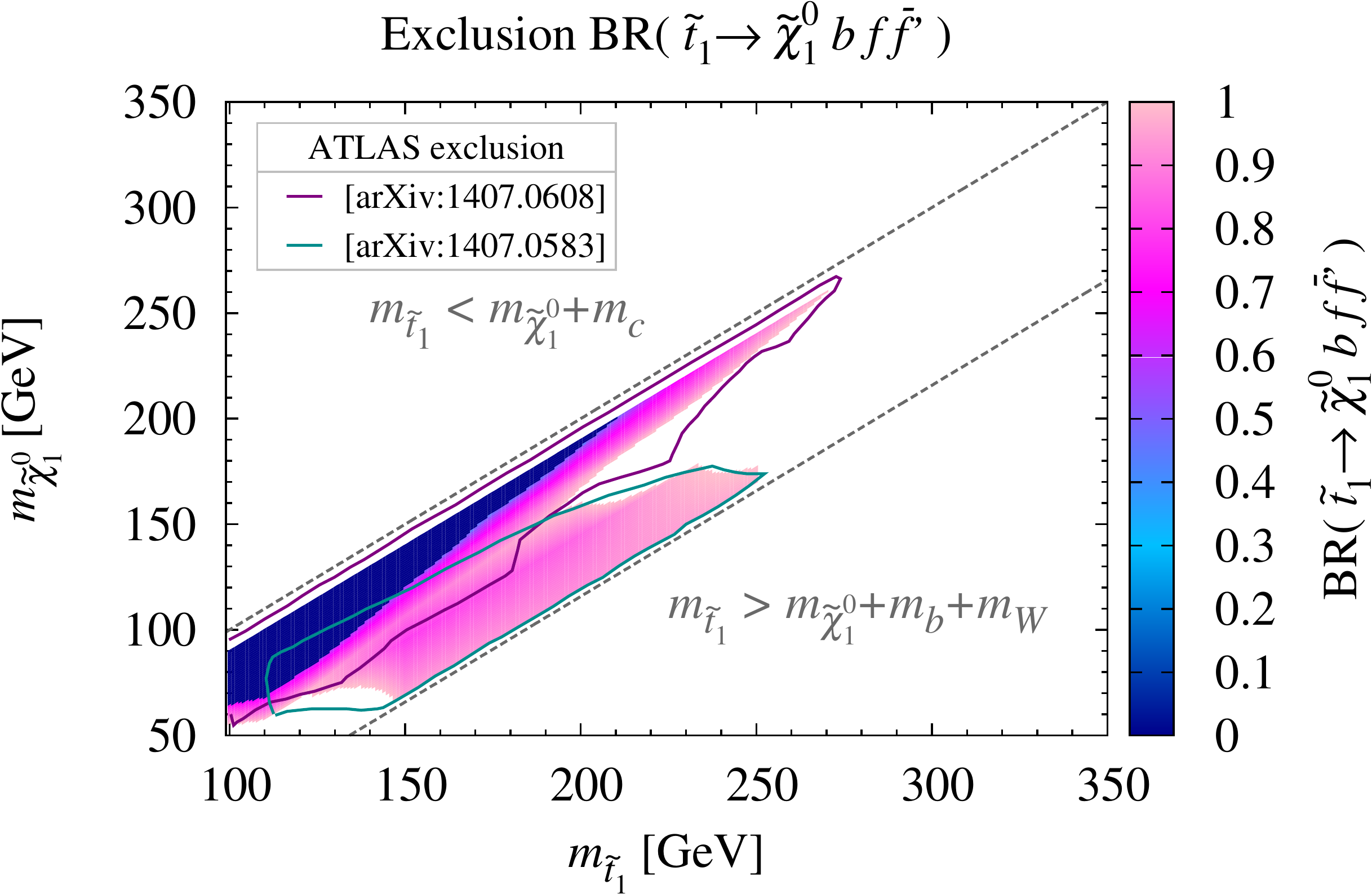} 
\caption{Exclusion limits in the
  $m_{\tilde{\chi}_1^0}-m_{\tilde{t}_1}$ plane at 95\% CL, based on
  the results for the $\tilde{t}_1 \to c \tilde{\chi}_1^0$ signature
  from \cite{Aad:2014nra} (left) and on the results for the $\tilde{t}_1
  \to \tilde{\chi}_1^0 b f\bar{f}'$ signature from
  \cite{Aad:2014nra,Aad:2014kra} (right).  The color code indicates the
  branching ratio down to which the exclusion limits are
  valid. Taken from \cite{Grober:2014aha}. \label{fig:stopexclusion}}   
\end{center}
\end{figure*}
%%%%%%%%%%%%%

For MFV scales large compared to the EWSB scale, the two-body decay
width is dominated by large logarithms of the ratios of these
scales, which should be resummed. Resummation effects can be included
by taking into account the FCNC couplings that are
induced through renormalization group running already at tree-level. This has been done in
Ref.\,\cite{Grober:2014aha} where we calculated the one-loop SUSY-QCD
corrections to the two-body decay. For the correct determination of
the $\tilde{t}_1$ branching ratios, also the four-body decay is
computed by consistently including FCNC couplings. Furthermore,
non-vanishing masses for the third generation fermions in the final
state have been taken into account. These decay widths have been implemented
in \textsc{Susy-Hit} \cite{susyhit,Djouadi:2006bz}. (The program with the newly
  implemented stop decays is available at \cite{program}.) The improved  $\tilde{t}_1$ branching ratios have implications for the LHC
stop searches and the bounds obtained on the mass of the lightest stop
$m_{\tilde{t}_1}$. To show this we have performed a
parameter scan and taken care to respect constraints from Higgs data,
SUSY searches, $B$-physics measurements 
and from the relic density. For details, we refer to
\cite{Grober:2014aha}. Figure~\ref{fig:stopexclusion} shows the
exclusion limits on $m_{\tilde{t}_1}$ depending on the mass
$m_{\tilde{\chi}_1^0}$, taking into account our results, i.e.\ as a function of the actual value of the $\tilde{t}_1$ branching
ratio. The plot reinterprets the ATLAS searches based on monojet-like
\cite{Aad:2014nra} and charm-tagged event selections
\cite{Aad:2014nra} and on searches for final states with one isolated
lepton, jets and missing transverse momentum \cite{Aad:2014kra}, where
limits on the lightest stop mass as a function of the neutralino mass
have been given, assuming a branching ratio of one.
The grey dashed lines in the figure limit the region in which 
%\begin{equation}
$
m_{\tilde{\chi}_1^0} + m_c \le m_{\tilde{t}_1} \le
m_{\tilde{\chi}_1^0} + m_b + m_W \;. 
$
%\label{eq:massinterval}
%\end{equation}
%
In this region the stop can be searched for in the
FV two-body decay and the four-body decay. 
The full pink line in the upper plot is the 95\% CL exclusion limit
based on combined charm-tagged and monojet 
ATLAS searches in the $\tilde{t}_1 \to c\tilde{\chi}_1^0$ decay
\cite{Aad:2014nra}, assuming a branching ratio of one. Under the
assumption that $\tilde{t}_1$ decays exclusively into the four-body
final state, ATLAS derived from the monojet analysis
\cite{Aad:2014nra} the exclusion given by the pink line (close to the upper
dashed line) in Fig.~\ref{fig:stopexclusion} (right) and from the
final states with one isolated lepton the exclusion region delineated by the green
line (close to the lower dashed line) \cite{Aad:2014kra}. With the
information given in \cite{Aad:2014nra,Aad:2014kra} we 
derived the exclusion limits for the two- and the four-body final
state as a function of the branching ratio, which is given by the
color code. 

The plot shows the stop masses that are excluded for a branching
ratio above the one associated with a specific color. Evidently, 
for smaller branching ratios the exclusion limits become
weaker. The combination of the two plots allows to extract the exclusion limits
for stops of a given mass as function of the neutralino mass and the
stop branching ratio. Thus it can be read off from
Fig.~\ref{fig:stopexclusion} (left) that $\tilde{t}_1$ masses of
150~GeV can be excluded for $m_{\tilde{\chi}_1^0}=80$~GeV if their branching
ratio into $c + \tilde{\chi}_1^0$ exceeds $0.43$. This in turn
implies that the stop four-body branching ratio is below 0.57. On
the other hand the right plot shows that in the same region stops can
be excluded if their branching ratio into the four-body final state
is larger than 0.88, which implies that the two-body decay branching
ratio is below 0.12 then. This means
that $m_{\tilde{t}_1} = 150$~GeV can be excluded for
$m_{\tilde{\chi}_1^0}=80$~GeV for scenarios in which $\mbox{BR}
(\tilde{t}_1 \to c \tilde{\chi}_1^0) < 0.12$ and $\mbox{BR}
(\tilde{t}_1 \to c \tilde{\chi}_1^0) > 0.43$, respectively, 
$\mbox{BR} (\tilde{t}_1 \to \tilde{\chi}_1^0 b f \bar{f}') >
0.88$ and $\mbox{BR} (\tilde{t}_1 \to \tilde{\chi}_1^0 b f \bar{f}') < 0.57$.
The dark blue regions correspond to stops with vanishing branching
ratios, so that all stop mass values associated with these regions are
excluded. In Fig.~\ref{fig:stopexclusion} (left) there is no smooth
transition between the dark blue and its neighbouring regions, as the exclusion
limits in the two-body final state are related to the ones in the
four-body final state which here apply for branching ratios $\gtrsim
0.46$, so that of course also in
Fig.~\ref{fig:stopexclusion} (right) there is no continuous color
gradient.

The exclusion limits given by the border of the colored
region at 100\% two-, respectively, four-body decay branching ratio, do not 
exactly match the ones derived by ATLAS. The reason is that ATLAS
provided information on the values of the excluded production cross
section times branching ratio only for a few points in the
$m_{\tilde{\chi}_1^0}-m_{\tilde{t}_1}$ plane so that a linear
interpolation between these points was necessary in order to cover the whole
region. Nevertheless, the agreement of the presented results with the given
exclusion limits is reasonably good. The advantage
of our approach is that it takes properly into account the information on the
actual stop branching ratios which can considerably weaken the stop
exclusion limits as is evident from Fig.~\ref{fig:stopexclusion}.  As
these plots can only be an approximation of what can be
done much more accurately by the experiments, they should be taken as
an encouragement to provide results also as function of the stop
branching ratios. 

%%%%%%%%%%%%%%%%%%%%%%%%%%%%%%%%%%%%%%%%%%%%%%%%%%%%%%%%%%%%%%%%
\section{Exclusive squark production and decay at the LHC}
\label{sec:production_and_decay}
%%%%%%%%%%%%%%%%%%%%%%%%%%%%%%%%%%%%%%%%%%%%%%%%%%%%%%%%%%%%%%%%

Fully differential NLO calculations of squark and gluino production
and decay have been presented recently ~\cite{Hollik:2012rc,
  Gavin:2013kga, Hollik:2013xwa,
  Gavin:2014yga, Boughezal:2012zb, Boughezal:2013pja}, including the matching to
parton showers for squark-squark and squark-antisquark
production~\cite{Gavin:2013kga, Gavin:2014yga}. These calculations show
that higher-order QCD effects may significantly modify the shape of
differential distributions. Thus, leading-order Monte-Carlo
predictions scaled with inclusive QCD corrections do, in general, not provide an accurate 
description of exclusive observables and cross sections with kinematic
cuts. We shall discuss the impact of the higher-order corrections on
the differential distributions following the analysis of
squark-antisquark and squark-squark production presented in
Ref.\,\cite{Gavin:2014yga}.

In order to quantify the impact of the parton shower on the
differential distributions, we have matched the NLO-calculation
presented in Section~\ref{sec:non_degenerate} with various parton
shower showers using the matching scheme of Ref.\,\cite{Nason:2004rx}
as implemented in the \PB~\cite{Alioli:2010xd}. The corresponding code is publicly
available and can be downloaded at \cite{spowheg}.  

Let us first comment on the impact of the NLO corrections on the shape of the differential distributions. 
We consider the production of a squark-antisquark pair, $pp \to \tilde{q}\bar{\tilde{q}}$, followed by (anti-)squark decay into 
a neutralino and a jet, $\sq/\bar{\tilde{q}}\rightarrow q/\bar{q}\, \neutone$. 
 In Fig.~\ref{fig:LONLOdec1} we present the LO and the NLO distributions for the transverse momentum of the hardest jet, $p_T^{j_1}$, using the benchmark scenario described in Section~\ref{sec:non_degenerate} (see Ref.\,\cite{Gavin:2014yga} for details). 
One observes a strong enhancement of the NLO corrections for small values of $p_T$, while they turn even negative for large values. The result for the second hardest jet, which is not shown here, is qualitatively the same. 
\bfig[t]
 \includegraphics[width=0.475\textwidth]{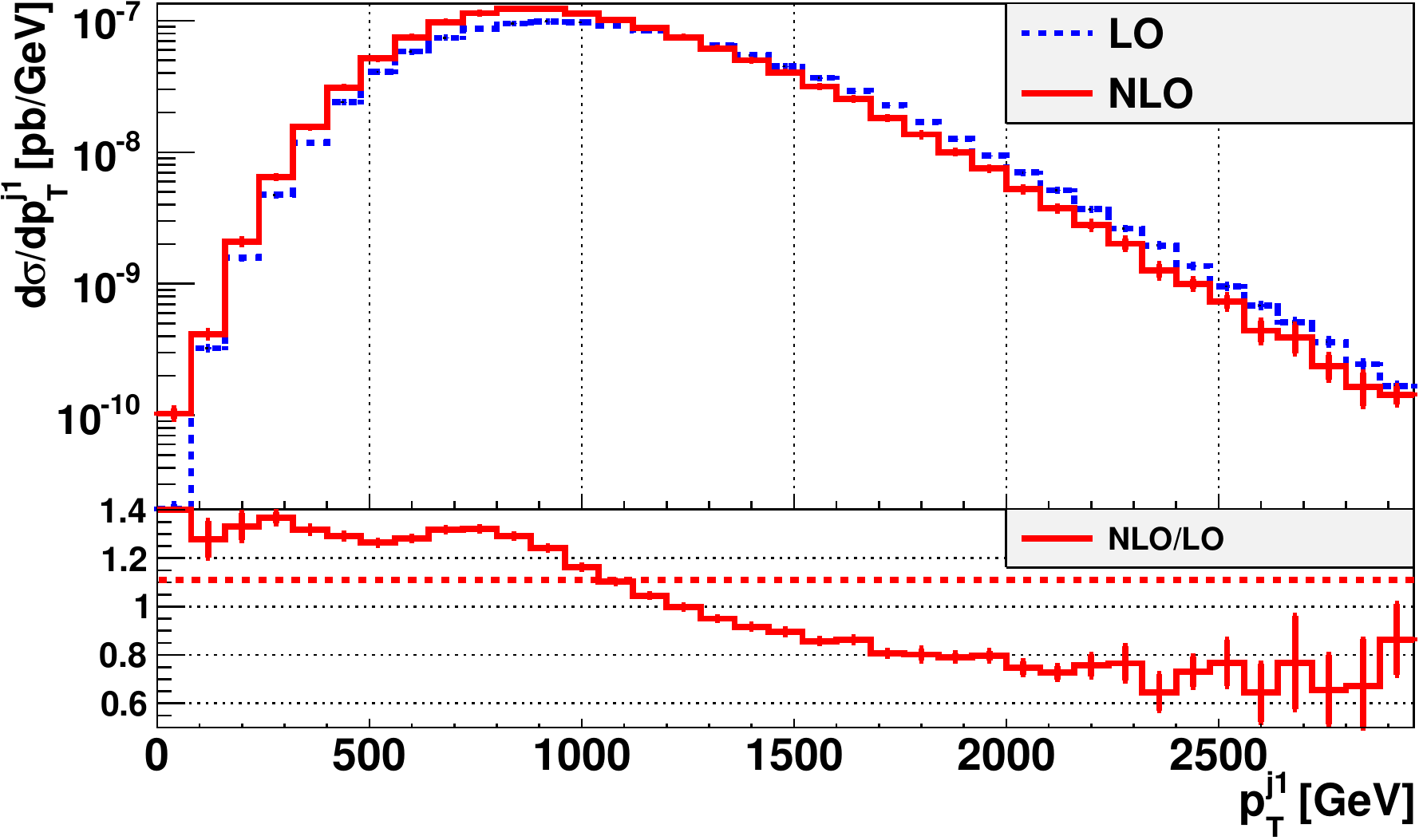}
\caption {\label{fig:LONLOdec1} Transverse momentum distribution of the hardest jet, $p_T^{j_1}$, for squark production at the LHC ($\sqrt{S}=14$\,TeV), combined with the subsequent (anti-)squark decay $\sq/\bar{\tilde{q}}\rightarrow q/\bar{q}\, \neutone$ for the SUSY scenario described in the text. Shown are the LO and NLO results, as well as the differential $K$-factor (full) and the total $K$-factor (dashed).}
\efig

More accurate predictions for differential distributions and exclusive observables require the matching of the NLO calculations with a parton shower. 
We have followed the matching scheme presented in Ref.\,\cite{Nason:2004rx} and implemented the NLO SUSY-QCD calculation 
of $pp \to (\tilde{q} \to q \neutone)(\bar{\tilde{q}}\to \bar{q}\neutone)+X$ into the \PB~\cite{Alioli:2010xd}. We have 
compared the default \textsc{Pythia\,6}~\cite{Sjostrand:2006za} and \textsc{Herwig++}~\cite{Bahr:2008pv} parton showers, and a $p_T$-ordered dipole shower~\cite{Platzer:2009jq, Platzer:2011bc}, which is also implemented in \textsc{Herwig++}. 
We find that the
predictions of the different parton showers for the observables
depending solely on the two hardest jets agree within $\matO(10\%)$ or
better. Comparing the showered results with the outcome of a pure NLO
simulation the effects of the parton showers on these observables are
at most of $\matO(20\%)$, except for the threshold region. Larger
deviations between the different parton showers emerge in the
predictions for the third hardest jet, which is formally described
only at LO in the hard process, see Fig.~\ref{fig:Shower1}. 

\bfig[t]
\includegraphics[width=0.465\textwidth]{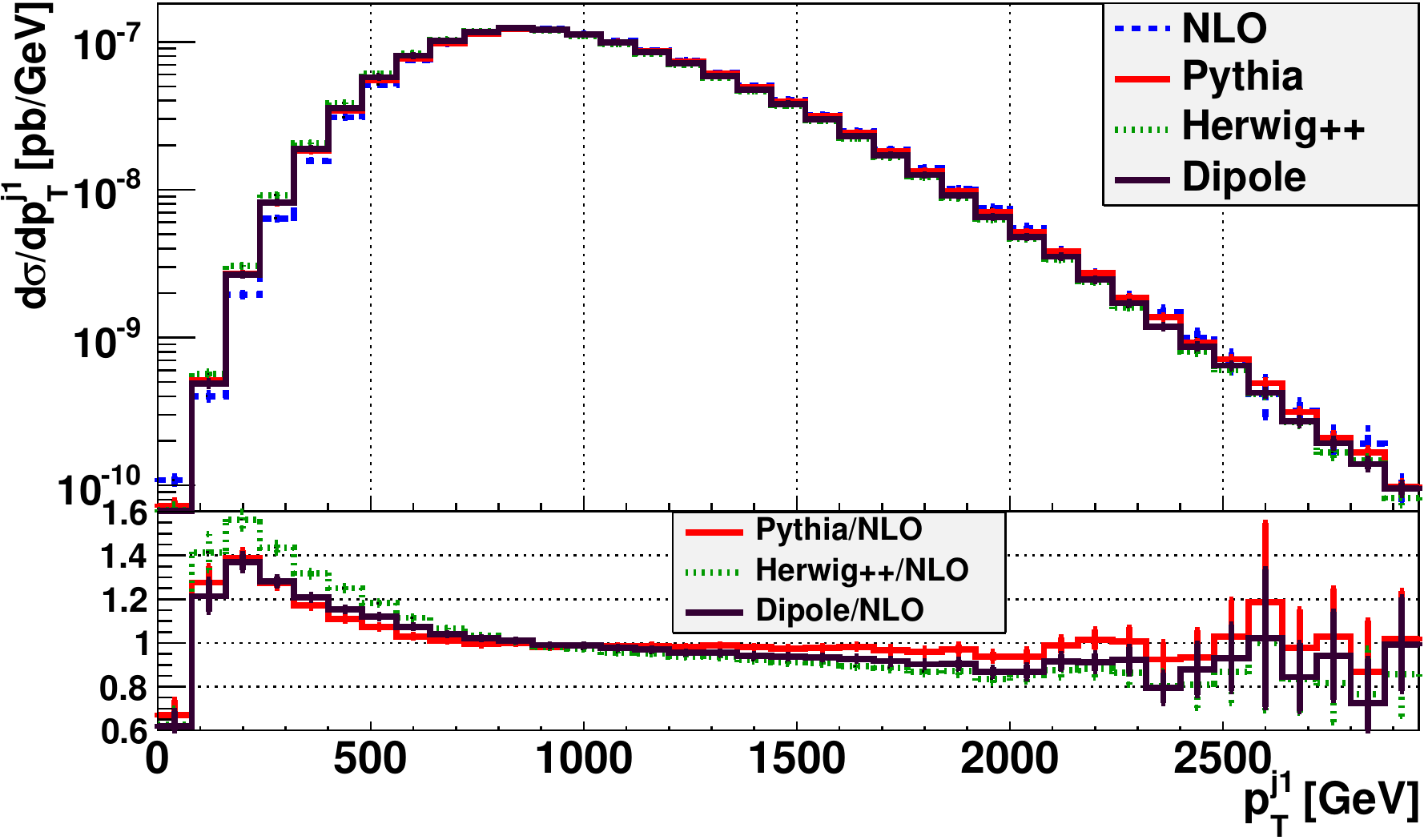}\\
\includegraphics[width=0.475\textwidth]{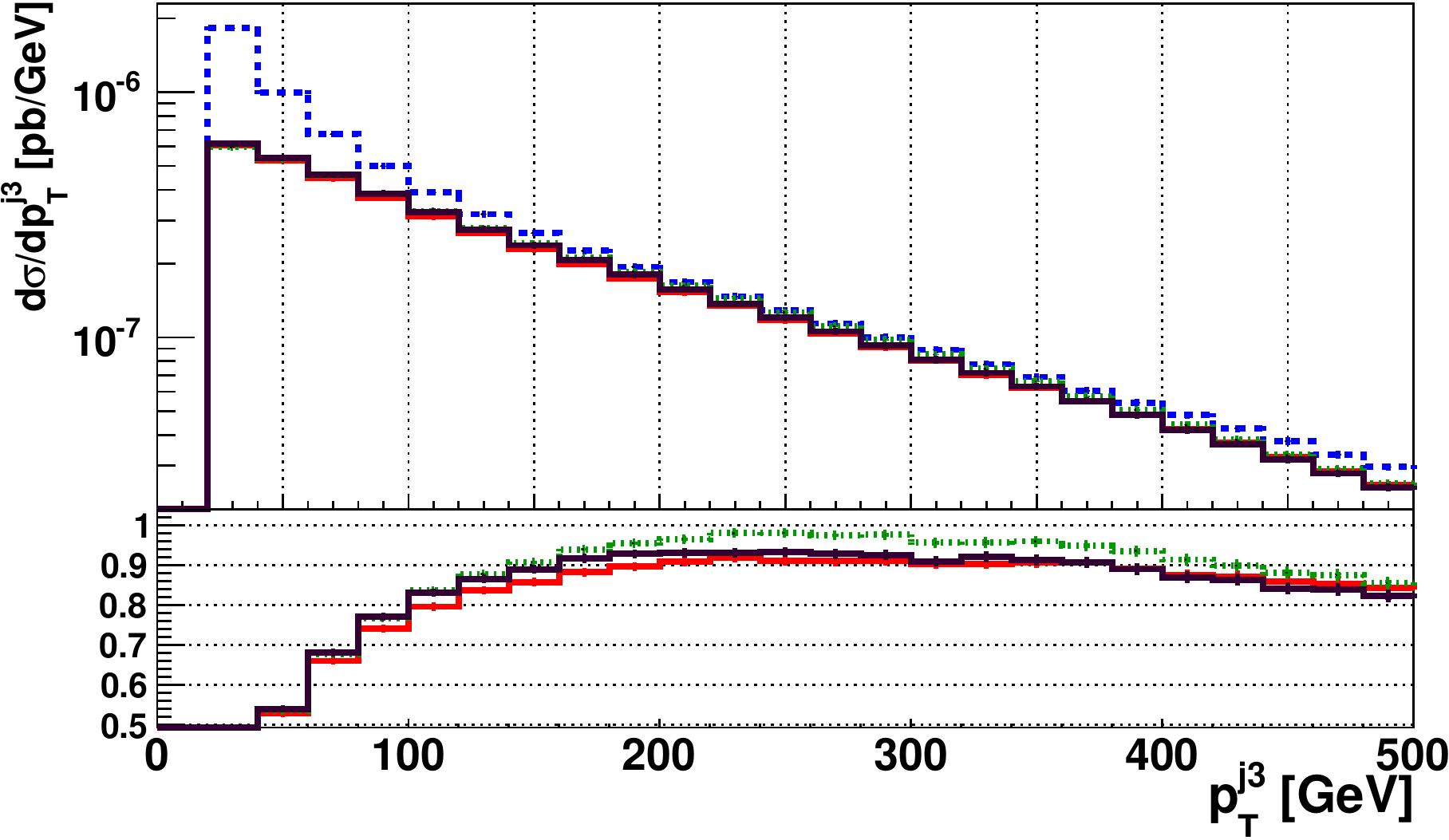}
\caption {\label{fig:Shower1} 
Transverse momentum distribution of the hardest (upper panel)  and third hardest (lower panel) jet, 
for squark production at the LHC ($\sqrt{S}=14$\,TeV), combined with the subsequent (anti-)squark decay 
$\sq/\bar{\tilde{q}}\rightarrow q/\bar{q}\, \neutone$ for the SUSY scenario described in the text. The NLO
  predictions and the NLO results matched to  the parton showers
  \PYTH~\cite{Sjostrand:2006za}, \HWG~\cite{Bahr:2008pv} and the
  \DS~\cite{Platzer:2009jq, Platzer:2011bc}, are shown. The lower
  inserts depict the respective ratios of the results obtained with the
  three parton showers and the pure NLO prediction.}
\efig

Let us finally comment on the impact of the higher-order corrections
on event rates after experimental selection cuts. We consider the
definition of a  signal region for the SUSY searches in
two-jet events performed by the ATLAS collaboration \cite{atlas_conf}
with event selection cuts corresponding to
\begin{gather}
  p_T^{j_1}>130\,\text{GeV}, \quad p_T^{j_2}>60\,\text{GeV},  \quad \slashed{E}_T>160\,\text{GeV}, \nonumber\\\
  \frac{\slashed{E}_T}{m_{\text{eff}}}>0.2,\quad m_{\text{eff}}^{\text{incl}}>1\,\text{TeV},  \quad \Delta\phi(j_{1/2},\vec{\slashed{E}}_T)>0.4,\nonumber\\
 \quad \mbox{and} \;\;\Delta\phi(j_3,\vec{\slashed{E}}_T)>0.4 \,\,\,\,\,\text{if}\,\,\,\,\, p_T^{j_3}>40\,\text{GeV}\,.
\label{eq:atlascuts}
\end{gather}
Here, the effective mass $m_{\text{eff}}$ is defined as the sum of the
$p_T$ of the two hardest jets and $\slashed{E}_T$, whereas the
inclusive definition of this observable includes all jets with
$p_T^j>40\,\text{GeV}$, \be m_{\text{eff}}^{\text{incl}} =
\sum_{i=1}^{n_j} p_T^{j_i} + \slashed{E}_T\,.  \ee Moreover,
$\Delta\phi(j_i,\vec{\slashed{E}}_T)$ denotes the minimal azimuthal
separation between the direction of the missing transverse energy,
$\vec{\slashed{E}}_T$, and the $i^{\text{th}}$ jet. The additional cut
$\Delta\phi(j_3,\vec{\slashed{E}}_T)>0.4$ is only applied if a third
jet with $p_T^{j_3}>40\,\text{GeV}$ is present.

\begin{table}
\renewcommand{\arraystretch}{1.25}
\bc
\begin{tabular}{c | c |c }
      & $\sq\sq$ & $\sq\sqbar$ \\\hline
NLO   & $0.871\,\text{fb}$  & $0.0781\,\text{fb}$  \\  
\PYTH & $0.883\,\text{fb}$  & $0.0797\,\text{fb}$  \\
\HWG  & $0.895\,\text{fb}$  & $0.0807\,\text{fb}$  \\
\end{tabular}
\caption{\label{tab:sigcuts} Total cross sections after applying the event selection cuts defined in Eq.~(\ref{eq:atlascuts}) for squark-squark and squark-antisquark production and subsequent decay $\sq\rightarrow q \neutone$ within the benchmark scenario introduced in Section~\ref{sec:non_degenerate} and specified in detail in Ref.\,\cite{Gavin:2014yga}. Results have been obtained at the level of a pure NLO simulation and including parton shower effects with \PYTH~and \HWG, respectively.}
  \ec
  \vspace*{-0.2cm}
\end{table}

Applying these cuts at the level of a pure NLO simulation yields the
event rates as shown in the first row of
Tab.~\ref{tab:sigcuts}. Results are shown for squark-squark and
squark-antisquark production and subsequent decay $\sq\rightarrow q
\neutone$ within the benchmark scenario introduced in Section~\ref{sec:non_degenerate}
and specified in detail in Ref.\,\cite{Gavin:2014yga}. Matching these
NLO results with a parton shower hardly affects the outcome after
using the cuts defined in Eq.~(\ref{eq:atlascuts}), as can be inferred
from the results obtained with \PYTH~and the \HWG~default shower
listed in the second and third row, respectively.

To conclude, we have discussed the effect of matching various parton showers to a NLO-QCD calculation of squark-squark and
squark-antisquark production. We find moderate shower effects of at most of $\matO(10-20\%)$ for the
observables depending solely on the two hardest jets, or for rather 
inclusive observables like event rates with selection cuts. In more exclusive distributions, 
including for example those for subleading jets, the effect of the parton shower may be 
more sizeable.  

%%%%%%%%%%%%%%%%%%%%%%%%%%%%%%%%%%%%%%%%%%%%%%%%%%%%%%%%%%%%%%%%
\section{Conclusions}
\label{sec:conclusions}
%%%%%%%%%%%%%%%%%%%%%%%%%%%%%%%%%%%%%%%%%%%%%%%%%%%%%%%%%%%%%%%%
Supersymmetric theories are among the most attractive extensions of the Standard Model, and searches for SUSY play 
a key role in the LHC physics program. A tremendous amount of work has been done in the last 20 years to calculate 
higher-order corrections to the production and decay of supersymmetric particles at hadron colliders. Such precision 
calculations are important to place limits on the masses of supersymmetric particles from current and future LHC searches, 
and they will be essential to determine the parameters of a supersymmetric theory in the case of discovery at the upcoming LHC run\,2. 

The production cross sections for the strongly interacting SUSY particles, squarks and gluinos, are known at next-to-leading order in SUSY-QCD, 
including the summation of threshold logarithms at next-to-next-to-leading accuracy. The uncertainty of the theoretical predictions 
is at a level of $\lesssim 15\%$ and dominated by the error on the parton distribution functions. Similarly, there is a vast literature 
on higher-order corrections to SUSY particle decay widths and branching ratios, covering most of the relevant decay modes in the MSSM.  

To fully exploit the potential of the LHC run\,2, it is crucial to obtain precise predictions for more exclusive observables, including 
differential distributions and cross sections with cuts. To this end, one needs to consider higher-order corrections to SUSY particle production and decay chains, and to match the NLO predictions with parton showers. First results of such a more comprehensive analysis have been presented here. 

Finally, future work should focus on developing automated tools to extend precision calculations to a wider class of beyond the SM 
physics scenarios.  
Such tools need to combine next-to-leading order accuracy in QCD for
generic beyond the SM processes with the summation of large logarithmic
threshold corrections and the matching with parton shower Monte Carlo programs. 

%%%%%%%%%%%%%%%%%%%%%%%%%%%%%%%%%%%%%%%%%%%%%%%%%%%%%%%%%%%%%%%%
\section*{Acknowledgements}
%%%%%%%%%%%%%%%%%%%%%%%%%%%%%%%%%%%%%%%%%%%%%%%%%%%%%%%%%%%%%%%%
We would like to thank our collaborators W.\ Beenakker, C.\ Borschensky, S.\ Brensing, T.\ van\ Daal, R.\ Gavin, R.\ Gr\"ober, C.\ Hangst, T.\ Janssen, 
M.\ Klasen, A.\ Kulesza, E.\ Laenen, R.\ van\ der\ Leeuw, S.\ Lepoeter, M.L.\ Mangano, L.\ Motyka, I.\ Niessen, S.\ Padhi, M.\ Pellen, T.\ Plehn, E.\ Popenda, X.\ Portell, M.\ Spira,  V.\ Theeuwes, A.\ Wlotzka, and P.M.\ Zerwas. 

This work was supported by the Deutsche Forschungsgemeinschaft through the collaborative research centre SFB-TR9 ``Computational Particle Physics", 
and by the U.S. Department of Energy under contract DE-AC02-76SF00515. 
MK is grateful to SLAC and Stanford University for their hospitality.

%\appendix

%\section{Special formule}

%% References

\nocite{*}
%\bibliographystyle{elsarticle-num}
%\bibliography{martin}



\end{document}